\documentstyle[12pt]{article}
\newcommand{\slp}{\raise.15ex\hbox{$/$}\kern-.57em\hbox{$\partial$}}
\newcommand{\sla}{\raise.15ex\hbox{$/$}\kern-.57em\hbox{$a$}}
\newcommand{\slA}{\raise.15ex\hbox{$/$}\kern-.57em\hbox{$A$}}
\newcommand{\slb}{\raise.15ex\hbox{$/$}\kern-.57em\hbox{$b$}}
\newcommand{\be}{\begin{equation}}
\newcommand{\tr}{{\rm tr}}
\newcommand{\ee}{\end{equation}}
\newcommand{\bear}{\begin{eqnarray}}
\newcommand{\ear}{\end{eqnarray}}
\newcommand{\ba}{\begin{eqnarray*}}
\newcommand{\ea}{\end{eqnarray*}}

\newcommand{\Aslash}{A\hskip-6.7pt /\hskip2pt}
\newcommand{\Dslash}{D\hskip-6.7pt /\hskip2pt}
\newcommand{\partialslash}{\partial\hskip-6.7pt /\hskip2pt}
\begin{document}


\begin{titlepage}
\setcounter{page}{1}
\begin{flushright}
HD--THEP--98--38\\
\end{flushright}
\vskip1.5cm
\begin{center}
{\large{\bf Thermal Bosonic and Fermionic Quantum Fields}}\\
{\large{\bf in Static Background Gauge Potentials}}\\
\vspace{1cm}
{A.A. Actor\footnote{email: aaa2@psu.edu} }\\
{\it Department of Physics}\\
{\it The Pennsylvania State University}\\
{\it Lehigh Valley Campus}\\
{\it Fogelsville, PA 18051, USA}
\\
K. D. Rothe \footnote{email: k.rothe@thphys.uni-heidelberg.de}
\\
{\it Institut  f\"ur Theoretische Physik - Universit\"at Heidelberg}
\\
{\it Philosophenweg 16, D-69120 Heidelberg, Germany}
\\
F.S. Scholtz \footnote{email: fgs@sunvax.sun.ac.za}
\\
{\it Department of Physics, University of Stellenbosch}
\\
{\it Stellenbosch 7600, South Africa}
\\

{(August 31, 1998)}
\end{center}

\begin{abstract}
We study at finite temperature the energy-momentum tensor
$T_{\mu\nu}(x)$ of (i) a scalar field in arbitrary dimension, and 
(ii) a spinor field in 1+1 dimensions, interacting with a static background  
electromagnetic field. $T_{\mu\nu}$
separates into an UV divergent part $T_{\mu\nu}^{sea}$
representing
the virtual sea, and an UV finite part $T_{\mu\nu}^{plasma}$
describing the thermal plasma
of the matter field. $T_{\mu\nu}^{sea}$
remains uniform in the presence of a \underbar{uniform}
electric field $\vec E$, while $T_{\mu\nu}^{plasma}$
becomes a periodic function
with period $\Delta x=2\pi T/E$
in the direction parallel to $\vec E$.
A related periodicity is found for a
uniform static magnetic field if one
spatial direction perpendicular to the magnetic field is compactified to a 
circle. 

\end{abstract}
\end{titlepage}

\section{Introduction}

A finite-temperature ($T>0$) or thermal quantum field can be visualized as 
a  
{\it sea} of virtual particles occupying space together with a thermal 
{\it  
gas}
of real field excitations. The virtual particle sea is independent of the  
temperature $T$. It can, however, be deformed by coupling
the field to static background structures of
various kinds. This is generally known as the static vacuum Casimir effect.
Somewhat less widely known is that, for
boundaries and other static backgrounds, the thermal gas is 
"mechanically"  
distorted along with the sea.

Abelian (and non-Abelian) gauge theories present
Casimir problems with particular features arising from the underlying gauge  
invariance. Indeed, the restrictions on the class of allowed
gauge transformations imposed by $T>0$ are found to have remarkable
consequences for the spatial energy distribution of a charged
thermal matter field  coupled to a static background electromagnetic 
field  
($\vec E, \vec B$). The local distortion by $(\vec E,\vec B)$
of the virtual sea and thermal {\it plasma}
(as we now refer to the thermal gas consisting of both particles and  
antiparticles)  can be revealed by a
local analysis in terms of the thermal stress energy momentum tensor.

The problem of charged quantum fields coupled to a uniform 
electromagnetic  
background field is an old one, going back to famous papers of Euler and  
Heisenberg \cite{Euler} and Schwinger \cite{Schwinger}.
There has been done a great deal of subsequent research,
much of it reviewed in  refs. \cite{Dittrich}-\cite{Sachs}.
However, global aspects of the problem have
received most of the attention, while local aspects seem to have been 
neglected.

In the present paper we try to gain
general insight into the local response
of gauged thermal matter fields to a background electromagnetic field.
Using the Matsubara formalism (see e.g. refs. \cite{Matsubara, Das, 
Rothe}),
we work on a hypercylinder of circumference $\beta = 1/T$ in the  
euclidean time direction,
choosing space to be flat and infinite. Our results reveal some
surprising features: We find for both scalar and spinor
fields that a uniform background electric field $\vec E$
causes the thermal plasma to become non-uniform -- in fact periodic in 
its  
spatial distribution along the direction of the $\vec E$,
with period $\Delta x = 2\pi T/{|E|}$. The sea nevertheless remains spatially 
uniform.

The organization of this paper is as follows: We begin by considering 
thermal  
scalar fields, for which the discussion can easily be carried out in 
arbitrary  
dimension. By coupling the scalar field
to an arbitrary static gauge potential $A_\mu(\vec x)$ we show that the  
characteristic effects arising from
the minimal coupling are common to all dimensions. Special attention is drawn to
 periodicity features related to gauge invariance
and to the topology
$S^1\times R^d$ of space-time. We then compute for the specific gauge  
potential $A_\mu = (Ex_1 + const,\vec 0)$ the thermal stress tensor  
$T^{(\beta)}_{\mu\nu}(x)$. The periodicity of the thermal plasma along $x_1$ with period $\Delta  x_1 = 2\pi T/E$ is thereby made explicit.

In section 3 we turn to the Schwinger model at $T>0$ with the
linear background potential referred to above. The discussion is now more  
involved, but the calculations reveal features similar to those  
encountered in the scalar case. We conclude in section 4 by 
commenting on a claim in the literature concerning the  factorization of the
thermal heat kernel, and drawing an analogy between our findings and 
the Quantum Hall Effect.
\section{Thermal scalar field}

Scalar electrodynamics is useful as a theoretical laboratory
for studying gauge
theory phenomena in arbitrary space-time dimension. We first consider
 the general problem of a thermal scalar field
$\hat\phi$ coupled to an arbitrary static 
i) background potential $V(\vec x)$ and 
ii) gauge potential $A_\mu(\vec x)$. 
We then specialize to the potential
$A_\mu=(Ex_1+const, \vec  0)$ for a uniform background electric
field $\vec E=(E,0,...,0)$ and compute explicitly the
thermal stress tensor of $\hat\phi$. The case of a uniform
 magnetic field is also discussed briefly.

\subsection{Scalar field in a static background
Schr\"odinger potential}

  To set the stage we briefly review the case of a scalar quantum 
field
interacting with a static background potential $V(\vec x)$
in $d$-dimensional free space $R^d$. We wish to study the thermodynamical
properties and vacuum Casimir energy of this system. To this end it will be convenient to work in the
imaginary time or Matsubara formalism. Euclidean space-time is then a
 hyper-cylinder $S^1\times R^d$. Correspondingly we impose 
periodic  
boundary conditions in euclidean time on the scalar  field $\phi(x_0,\vec 
x)$  
\footnote{Throughout this section we use the euclidean notation $x=
\{x_\mu\} =  (x_0,\vec x)$.},
\be\label{2a}
 \phi(x_0,\vec x) =  \phi(x_0 + \beta,\vec x)
\ee
where $\beta = 1/T$ and $T$ is the temperature. The spectral
operator for the theory in question is $[-\partial_0^2 -\Delta + V(\vec 
x)]$  
with $\Delta$ the Laplacian in $d$ dimensions. The vacuum and
thermodynamical properties of the system can be computed from the bilocal
heat kernel
\[
h^{(\beta)}(t;x,y) = \sum_{k} 
e^{-t\lambda^2_{k}}\phi_{k}(x)\phi^*_{k}(y)\;,
\]
where $\lambda^2_{k}$ and $\phi_k(x_0,\vec x)$ are the eigenvalues and
respective eigenfunctions of
the spectral operator,
\[
[-\partial_0^2 - \Delta + V(\vec x)]\phi_k(x) = \lambda_k^2 \phi_k(x)\;.
\]
With $\phi(x)$ subject to the boundary condition (\ref{2a})
we have
\bear\label{2b}
\phi_k(x) \to \phi_{mn}(x) &=& \frac{1}{\sqrt\beta}e^{i(\frac{2\pi
m}{\beta})x_0}\varphi_n(\vec x)\;,\nonumber\\
\lambda^2_k \to \lambda^2_{mn} &=& (\frac{2\pi m}{\beta})^2 + \omega_n^2\;,
\ear
where the spatial modes $\varphi_n(\vec x)$ and associated
spectum $\{\omega_n^2\}$ are determined by the spatial
mode equation
\[
[-\Delta + V(\vec x)]\varphi_n(\vec x) = \omega^2_n \varphi_n(\vec x)\;.
\]
The Euclidean thermal Green function \footnote{We denote operators by a
"hat".} then has the spectral representation
\ba
<\hat\phi(x) \hat\phi(y)>_\beta &=&
\sum_{m,n} \frac{\phi_{mn}(x)\phi_{mn}(y)^*}{[(\frac{2\pi m}{\beta})^2 +
\omega_n^2]}\\
&=& \int_0^\infty dt \frac{1}{\beta}\sum_{m} e^{-t(\frac{2\pi m}{\beta})^2}
e^{i\frac{2\pi m}{\beta}(x_0 - y_0)}
\sum_n e^{-t\omega^2_n}\varphi_n(\vec x)\varphi_n^*(\vec y).\nonumber
\ea
We may perform  the Matsubara sum by using the Jacobi identity
\cite{Bateman}
\bear\label{2c}
\sum_{m=-\infty}^{\infty}e^{-b(m-a)^2}=\sqrt{\frac{\pi}{b}}
\sum_{l=-\infty}^\infty e^{-\frac{\pi^2 l^2}{b}}e^{-i2\pi a l}
\ear
with the result
\bear\label{2d}
<\hat\phi(x) \hat\phi(y)>_\beta &=& \int_0^\infty dt
h(t;x,y)_{T=0}\nonumber\\
&\times&\sum_{l=-\infty}^{\infty} e^{-\frac{l^2\beta^2}{4t}}e^{\frac{l\beta}{2t}(x_0-y_0)}\ear
where
\bear\label{2e}
h(t;x,y)_{T=0}&=& \sqrt{\frac{1}{4\pi t}}e^{-
\frac{(x_0 - y_0)^2}{4t}}\times\nonumber\\
&\times&\sum_{n} e^{-t\omega_n^2} \varphi_n(\vec x)\varphi_n^*(\vec y)
\ear
is the $T=0$ bilocal heat kernel of the operator
$[-\partial_0^2 - \Delta + V(\vec x)]$.
Hence, for a static background potential the $T>0$ Green function
separates  \cite{Actor} into
two distinct and well defined parts: the {\it virtual sea} part ($l=0$
contribution)
which is independent of $T$ 
and coincides with the $T=0$ Green function,
and the {\it thermal gas} part ($l \neq 0$ contribution) which exhibits the
full temperature dependence and vanishes exponentially as $T\rightarrow 0$:
\[
<\hat\phi(x) \hat\phi(y)>=<\hat\phi(x) \hat\phi(y)>_{sea}
+ <\hat\phi(x) \hat\phi(y)>_{gas}\;.
\]

Now suppose that we have calculated from $<\hat\phi(x)\hat\phi(y)>_{sea}$
the $T=0$ vacuum stress tensor
\be\label{2f}
T^{\mu\nu}_{sea}\equiv <\hat T^{\mu\nu}>=
\sum_nT_n^{\mu\nu}\ee
as a (still to be renormalized) spatial mode sum. Repeating the
calculation at finite temperature we obtain from eq. (\ref{2d})
\bear\label{2ff}
T^{(\beta)\mu\nu}&=&<\hat T^{\mu\nu}>_\beta\nonumber\\
 &=& \sum T^{\mu\nu}_n
\frac{1 + e^{-\beta \omega_n}}{1 - e^{-\beta \omega_n}}\nonumber\\
&=& T^{\mu\nu}_{sea} + \sum_n T^{\mu\nu}_n \frac{2}{e^{\beta 
\omega_n}-1}\nonumber\\
&=&T^{\mu\nu}_{sea} + T^{\mu\nu}_{gas} \;.\
\ear
At finite temperature the mode sum is thus modified by the
familiar Bose-Einstein distribution, in agreement with one's
expectations. This completes our brief review of a usual
Casimir problem at $T>0$. 


\subsection{Scalar field in a static background gauge potential }

Let us now consider a (massive or massless) scalar quantum field $\phi(x)$
coupled to a {\it static} background
Abelian gauge potential $A_{\mu}(\vec x)$. We shall again be interested in
studying the thermodynamical properties and Casimir energy of this system.

On a space-time hyper-cylinder $S^1\times R^d$,
both $\phi(x_0,\vec x)$ and
$A_\mu(x_0,\vec x)$ are required to satisfy  periodic boundary conditions
in euclidean time:
\be\label{2g}
 \phi(x_0,\vec x) =  \phi(x_0 + \beta,\vec x)\;\;,
A_\mu(x_0,\vec x) = A_\mu(x_0 + \beta,\vec x)
\ee
The relevant spectral operator in this case is the
gauged Laplacian in $d+1$ dimensions $-{ D}^2_\mu$, where
$D_\mu = \partial_\mu -iA_\mu$ couples the quantum scalar field to a static
background gauge potential $A_\mu(\vec x)$. (We have absorbed the electric
charge into $A_\mu$.)
The thermodynamical and vacuum
properties of the system can again be computed from
the bilocal heat kernel
\[
h^{(\beta)}(t;x,y) = \sum_{k} e^{-t\lambda^2_k}e^{-tM^2}
\phi_{k}(x)\phi_{k}^*(y)
\]
where $\lambda^2_{k}$ are the eigenvalues of  $-{D}^2_\mu$, $M$ is the mass 
and $\phi(x)$ is
subject to the boundary condition (\ref{2g}).
Periodicity in $x_0$ implies
\[
\phi_k(x) \to \phi_{mn} =
\frac{1}{\sqrt\beta}e^{i\frac{2\pi m}{\beta}x_0}\varphi_{mn}(\vec x)\;,
\]
where $\varphi_{mn}(\vec x)$ now satisfies the associated eigenvalue 
problem
\be\label{2h}
\left[ -(\vec\bigtriangledown - i\vec A(\vec x))^2 + V_m(\vec x)\right]
\varphi_{mn}(\vec x) = \lambda^2_{mn}\varphi_{mn}
\ee
with
\be\label{2i}
V_m(\vec x) = [A_0(\vec x) - \frac{2\pi m}{\beta}]^2\;.
\ee
Notice that the $m$-dependence of the Schr\"odinger-like
background potential $V_m(\vec x)$  leads to a coupling of  spatial
position $\vec x$ with the Matsubara frequencies. Eqs. (\ref{2h}) thus
represent a different equation for each Matsubara frequency $\frac{2\pi
m}{\beta}$, with $n$ labelling the complete set of normalizable solutions
$\{\varphi_{mn}\}$ of this Schr\"odinger problem for a given potential
$V_m(\vec x)$.  The situation is thus very different from the
scalar case discussed previously. It is characteristic of gauge theories
and has very important consequences as we shall see.

On the cylinder $S^1\times R^{d}$  we can always gauge a static 
$A_0(\vec x)$ to the interval $[0,\frac{2\pi}{\beta}]$,
but not in general to zero, if we respect the periodicity property (\ref{2g})
of $A_\mu$ and $\phi$.
The only exception is when $A_0 = N\frac{2\pi}{\beta}$,
in which case we may gauge $A_0$
to zero by performing the (allowed) gauge transformation 
$A_0 \rightarrow A_0  +\partial_0\lambda$ with $\lambda = (\frac{2\pi N}{\beta})x_0$.

Another way of stating this is to observe
that in the exceptional case $A_0 = N\frac{2\pi}{\beta}$ the
gauge transformation can be absorbed into the Matsubara index $m$ via the
transformation $m\rightarrow m+N$. For this reason $A_0(\vec x)$ is always
gauge equivalent to a configuration taking values in the range
$[0,\frac{2\pi}{\beta}]$. In the
zero-temperature limit, on the other hand, we
may always gauge $A_0(\vec x)$ to zero. Indeed, the discrete Matsubara
frequencies $k_0 = \frac{2\pi m}{\beta}$
become a continuous variable $k_0$
in the range $-\infty < k_0 < \infty$, and
$A_0(\vec x)$ may be absorbed into a shift in $k_0$
under the integral $\int dk_0$.

\bigskip
\noindent{\it Green function}

\bigskip
Following the steps of the previous section, we have this time for the
thermal Green function
\bear\label{2j}
<\hat\phi(x) \hat\phi(y)>_\beta &=& \sum_{m,n}
\frac{\phi_{mn}(x)\phi_{mn}^*(y)}{[\lambda_{mn}^2 + M^2]}\nonumber\\
&=&\int_0^\infty dt\;e^{-tM^2}\frac{1}{\beta}
\sum_{m}e^{i\frac{2\pi m}{\beta}(x_0 - y_0)} h_m(t;\vec x,
\vec y)
\ear
where
\be\label{2k}
h_m(t;\vec x,\vec y) =
\sum_n e^{-t\lambda^2_{mn}}\varphi_{mn}(\vec x)\varphi_{mn}^*(\vec y)
\ee
is the spatial bilocal heat kernel for the $m$'th spatial background
potential $V_m(\vec x)$ in (\ref{2i}) . Notice that because of the coupling 
of Matsubara frequencies with the spatial modes the Green function can
no longer be straightforwardly separated into ``sea'' and ``gas''
contributions as in the potential problem discussed previously.
In special cases
this separation can nevertheless be made explicit, as we shall see.

\bigskip
\noindent {\it Energy-momentum tensor}

\bigskip
The symmetric canonical energy-momentum tensor for the complex scalar field
$\hat\phi(x)$ coupled to a background $A_\mu$ is formally given by
(in Minkowski space-time)
\[
\hat T_{\mu\nu} = \frac{1}{2}[(D_\mu\hat\phi)^\dagger (D_\nu\hat\phi) +
(D_\nu\hat\phi)^\dagger (D_\mu\hat\phi)] - \eta_{\mu\nu}\hat{\cal L}
\]
where
\[
\hat{\cal L} = \frac{1}{2}[(D_\mu\hat\phi)^\dagger (D^\mu\hat\phi)
- M^2 \hat\phi \hat\phi^\dagger].
\]
Using the equation of motion $(D_\mu D^\mu + M^)\hat\phi= 0$ we have
for the divergence of $\hat T_{\mu\nu}$,
\[
\partial^\mu \hat T_{\mu\nu} = F_{\mu\nu} \frac{i}{2}[\hat\phi^\dagger 
(D^\mu  
\hat\phi) - (D^\mu\hat\phi)^\dagger \hat\phi]
\]
and for the trace,
\be\label{2l}
\hat T_\mu^\mu = -\frac{1}{2}(d-1)(D_\mu\hat\phi)^\dagger (D^\mu\hat\phi)
+ \frac{1}{2}(d+1)M^2\hat\phi^\dagger \hat\phi \;.
\ee
From (\ref{2l}) we see that $\hat T_{\mu\nu}$ is traceless only in $d=1$
spatial dimension for $M=0$. We furthermore see that $\hat T_{\mu 0}$ is
conserved if $F_{i0}=0$, that is, if the electric field vanishes. This
still allows for a static magnetic field, which makes sense, since a static
magnetic field cannot do work on charges.

From (\ref{2d}) the thermal stress energy tensor $T^{(\beta)}_{\mu\nu} =
<\hat T_{\mu\nu}>_\beta$ is now easily written down.
We have for the separate
components
\bear\label{2m}
<D_0\hat\phi(x)[D_0\hat\phi(x)]^\dagger>_\beta 
&=& \int_0^\infty dt\;e^{-tM^2}\frac{1}{\beta}(\frac{2\pi}{\beta})^2
\sum_m [m - a(\vec x)]^2 h_m(t;\vec x,\vec y)\nonumber\\
\ear
\bear\label{2n}
<D_i\hat\phi(x)[D_j\hat\phi(x)]^\dagger>_\beta
&=& \int_0^\infty dt\;e^{-tM^2}\frac{1}{\beta}
\sum_m \lim_{{\vec x}\to {\vec y}}\{(D^x_i)^\dagger D^y_j
h_m(t;\vec x, \vec y)\}\nonumber\\
\ear
\bear\label{2o}
<D_0\hat\phi(x)[D_j\hat\phi(x)]^\dagger>_\beta 
&=& \int_0^\infty dt\;e^{-tM^2}\frac{1}{\beta}(\frac{2i\pi}{\beta})
\sum_m [m - a(\vec x)]\nonumber\\
&\times& \lim_{{\vec x}\to {\vec y}}\{(D^y_j)^\dagger
h_m(t;\vec x, \vec y)\}
\ear
\be\label{2p}
  <|\hat\phi(x)|^2>_\beta = \int_0^\infty dt\;e^{-tM^2}\frac{1}{\beta}
\sum_mh_m(t;\vec x,\vec x)  \ee
where
\[ a(\vec x) = \frac{\beta}{2\pi}A_0(\vec x) \]
is the rescaled temporal component $A_0$.

The above expressions giving the energy-momentum tensor in terms
of the heat kernel  need of course to be properly UV-regularized.
We see that these  results differ essentially from the nongauge
results given earlier, since
in the present case $T_{\mu\nu}$ is expressed
as a Matsubara sum over \underbar{nonidentical}
spatial problems, the latter depending on
Matsubara label $m$. Thus an explicit
separation into a  sea and thermal gas (or more properly, thermal plasma)
contributions is in general not possible. We now consider specific 
background  
potentials for which the problem is completely soluble, and the
separation into  gas and thermal plasma contributions can be displayed.
\subsubsection{Constant gauge potential}

To begin with we consider a constant background gauge potential
\be\label{2q}
  A_0 = \frac{2\pi}{\beta}a,\quad \vec A=0.\ee
As already pointed out, on the cylinder
$S^1\times R^d$ this gauge potential
cannot be gauged to zero; however, it is
gauge equivalent to a potential with
$a$ in the range $0\leq a \leq 1$. On the other hand, the spatial
component $\vec A$ of a constant $A_\mu$
can always be gauged to zero on this
cylinder, so that we may choose $\vec A = 0$.

Following our general notation  we have in this case (for infinite volume the index $n$
becomes the  continuous momentum label $\vec k$)
\bear\label{2r}
\phi_{mk}(\vec x) &=& \frac{1}{(2\pi)^{\frac{d}{2}}}e^{i\vec k\cdot \vec
x}\nonumber\\
 \lambda^2_{mk} &=& (\frac{2\pi}{\beta})^2 (m-a)^2 + {\vec k}^2\ear
and
\bear\label{2s}
V_m(\vec x) &=& (\frac{2\pi}{\beta})^2 (m-a)^2 \nonumber\\
h_m(t;\vec x,\vec y) &=& e^{-(\frac{2\pi}{\beta})^2 (m-a)^2 }
h_0(t;\vec x-\vec y)
\ear
where
\ba\label{2t}
h_0(t;\vec x-\vec y) &=& \frac{1}{(2\pi)^d} \int d^dk \;e^{-t{\vec k}^2}
e^{i\vec k\cdot(\vec x-\vec y)}\nonumber\\
&=& \frac{1}{(4\pi t)^{\frac{d}{2}}}e^{-\frac{(\vec x - \vec y)^2 }{ 4t}}
\ea
is the infinite volume, zero temperature heat kernel of the free scalar field.

The factorization of the heat kernel $h_m$ into an $m$-dependent and
and $m$-independent factor  now enables one to perform  Matsubara sums
explicitly by making use of the identity
\bear\label{2u}
\sum_{-\infty}^{\infty} (m-a)^2 e^{-b(m-a)^2}
&=&\frac{1}{2b}\sum_{-\infty}^{\infty} e^{-b(m-a)^2}\nonumber\\
&-& 2(\frac{\pi}{b})^{\frac{5}{2}}\sum_{n=1}^{\infty}n^2
e^{-\frac{n^2\pi^2}{b}} \cos(2\pi an)
\ear
obtained from (\ref{2c}) by differentiation with respect to $b$.
One finds for the Minkowskian \footnote{Note that the transition from
Euclidean to Minkowskian space-time involves a minus
sign:$<-|D_0\hat\phi|^2>_\beta \to <|D_0\hat\phi|^2>_\beta\;$.} energy 
density  
after some simple algebra
\be\label{2v}
T^{(\beta)}_{00} = T^{sea}_{00}+T_{00}^{plasma}\ee
where the temperature-independent part representing the sea is given by the
UV-divergent integral
\be\label{2w}
T_{00}^{sea} = \frac{1}{2}\int_0^\infty dt\;
e^{-tM^2}(4\pi t)^{-(\frac{d+1}{2})}[\frac{d-1}{2t} + M^2]\;,
\ee
and the temperature dependent part representing the thermal
plasma  carries all the dependence on the gauge potential and is finite:
\bear\label{2x}
T_{00}^{plasma}  &=& \int_0^\infty dt\;
e^{-tM^2}(4\pi t)^{-(\frac{d+1}{2})}\sum_{n=1}^{\infty}
\cos(2\pi an)\nonumber\\
&\times&\{ \frac{n^2\beta^2}{4t^2} + \frac{d-1}{2t} + M^2\}
e^{-\frac{n^2\beta^2}{4t}}.
\ear
The integral in (\ref{2x}) can  be evaluated in terms of the
modified Bessel function $K_\nu(z)$, but we shall not do so.
It is important to observe that the thermal part vanishes exponentially as
$T\to 0$, and that it exhibits the expected periodicity property in the
parameter $a$, in line with our earlier observation that $a$ can always be
chosen to lie in the interval $[0,1]$.

The remaining diagonal stress tensor elements $T_{ii}(\beta)$ with
$i=1,2,..,d$ are given by $T_{00}(\beta)$ with the signs of the
$(d-1)/2t$ and $M^2$ terms in the curly brackets of eqs. (\ref{2w}),
(\ref{2x}) reversed. All nondiagonal elements of $T_{\mu\nu}^{(\beta)}$
vanish. This tensor is obviously conserved.

Finally observe that for $d=1$ and $M=0$ we have $T_{\mu\nu}^{sea} = 0$, 
and  
there is no need to perform an UV renormalization.
Furthermore $T_{00}^{plasma} = T_{11}^{plasma}$ in this case, in accordance
with the trace condition (\ref{2h}).
\subsubsection{ Constant electric field}
Next we consider the linear background potential
\be\label{2y}
A_0(x_1)=Ex_1+2\pi a/\beta,\quad \vec A=0\ee
corresponding to a constant background electric field $\vec {\cal E}=
(-E,0,...,0)$ in the $x_1$ direction. Here the constant term
$2\pi a/\beta$ in $A_0$ has the obvious physical significance of
determining the location $x_1=-2\pi a/\beta E$ of the zero
of this linear potential. Global aspects of what now follows
have been investigated previously for scalar fields
as well as for spinor fields (see e.g. refs. \cite{Dittrich}-\cite
{Sachs}).
We focus here  on local aspects of this problem. Note that the limit
$E\to 0$ is distinctly nontrivial. For that reason we have
chosen to analyse separately the $E\not=0$ problem here
and the $E=0$ problem in subsection 2.2.1 above.

Continuing with our general notation (where now  $n\to
(n,\vec k_\perp))$ we have the spatial modes
\be\label{2z}
\varphi_{mn\vec k_\perp}(\vec x)=\varphi_n(x_m)(2\pi)^{-\frac{1}
{2}(d-1)}
e^{i\vec k_\perp\cdot\vec x_\perp}\ee
with $\vec k_\perp=(k_2,...,k_d)$ and $\vec x_\perp=(x_2,...,x_d)$
representing momentum and position perpendicular to $x_1$. Inserting
$\varphi_{mn\vec k_\perp}$ into the spatial mode
equation (\ref{2h}) one obtains
\be\label{2aa}
\left[-\frac{d^2}{dx^2_m}+E^2x^2_m\right]\varphi_n(x_m)
=\epsilon_n\varphi_n(x_m),\ee
where
\be\label{2ab}
x_m\equiv x_1+\frac{2\pi}{\beta E}(a-m).\ee
This is just the harmonic oscillator eigenvalue problem in
Schr\"odinger theory with orthonormal eigenfunctions
\bear\label{2ac}
\varphi_n(x_m)&=&2^{-n/2}\frac{1}{\sqrt{n!}}\left(
\frac{E}{\pi}\right)^{\frac{1}{4}}e^{-\frac{1}{2}Ex^2_m}
H_n(\sqrt E x_m),\nonumber\\
\epsilon_n&=& 2E(n+\frac{1}{2}),\quad n=0,1,2,...\ear
Here $H_n(z)$ are Hermite polynomials satisfying
$y''-2zy'+2ny=0$. In eq. (\ref{2h})
the $m$-dependent backgrounds $V_m(x_1)=E^2x_m^2$ are identical harmonic
oscillator potentials centered at equidistant positions
$x_1=(m-a)2\pi/\beta E$. As we shall see this periodic
arrangement of identical potentials leads to a periodic structure along
$x_1$ (with period $\Delta x_1=2\pi/\beta E)$ in $T_{\mu\nu}(x)$
and in other local quantum functions. Notice that the $m$ dependence of
$\varphi_{mn\vec k_\perp}(\vec x)$ resides entirely in
the argument $x_m$ of the harmonic oscillator wave function
$\varphi_n(x_m)$ -- i.e. entirely in the position $x_1=
(m-a)2\pi/\beta E$ of the zero of $V_m(x_1)$. One consequence
of this fact is that the spectrum of $-D^2$ given by
\be\label{2ad}
\lambda^2_{mnk_\perp}=2E(n+\frac{1}{2})+\vec k_\perp^2\ee
does \underline{not} depend on the Matsubara label $m$.
This is in sharp contrast with the nongauge scalar theory
of section 2.1 with its spectrum (\ref{2b}), and with the
constant $A_\mu$ problem above with spectrum (\ref{2r}).

The spatial heat kernel (\ref{2k}) constructed from the spatial
modes (\ref{2z}) is
\be\label{2ae}
h_m(t;\vec x,\vec y)=k_E(t;x_m,y_m)h_0(t;|\vec x_\perp-
\vec y_\perp |)\ee
where
\bear\label{2af}
&&k_E(t;x_m,y_m)=\sum^\infty_{n=0}e^{-t\lambda^2_n}
\varphi_n(x_m)\varphi_n^*(y_m)\nonumber\\
&&=\left[\frac{E}{2\pi \sinh (2Et)}\right]^{\frac{1}{2}}
e^{-\frac{1}{2}E(x_1-y_1)^2\coth(2Et)}\nonumber\\
&&\times e^{-Ex_my_m\tanh(Et)}\ear
and $h_0$ is the free-space heat kernel  for $d-1$
dimensions. The mode sum $\sum_n$ here has been
performed with the help of the identity (see e. g. ref. \cite{
Bateman}, p. 194)
\[\sum^\infty_{n=0}\frac{1}{n!}\left(\frac{z}{2}\right)^n
H_n(x)H_n(y)=(1-z^2)^{-1/2}\exp\left\{
\frac{z}{1-z^2}[2xy(1-z)-z(x-y)^2]\right\}.\]
Alternatively one can use known formulae for the
propagator in the harmonic oscillator problem (see section 3). Note that in the
limit $E\to 0$ the heat kernel (\ref{2af}) smoothly becomes the
one-dimensional free-space heat kernel as it should
\[k_E(t;x_m,y_m)\to\frac{1}{\sqrt{4\pi}}
e^{-(x_1-y_1)^2/4t},\quad E\to 0.\]

The limit $E\to 0$ is nonetheless far from being uniform:
As $E\to 0$ the potentials $V_m(x_1)=E^2x^2_m\to(2\pi m/\beta)^2$
change into $m$-dependent constants (like mass terms)
independent of $x_1$.
The background returns to the constant potential
$A_\mu=(2\pi a/\beta, \vec 0)$ of the preceeding
subsection with its spectrum $\lambda^2_{mk}=(m-a)^2(2\pi/\beta)^2+
\vec k^2$. Remarkably, the dependence on $m$ so conspicuously absent
from the $E>0$ spectrum $\lambda^2_{mnk_\perp}$ reenters the $E=0$
spectrum.

The Green function (\ref{2j}) and local quantities derived from
it possess the sea + plasma structure one expects to find. Let
us display this for the Green function (\ref{2j}) in the limit
$y\to x$;
\bear\label{2ag}
<|\hat\phi(x)|^2>_\beta&=&\int^\infty_0dt e^{-tM^2}(4\pi t)^{-d/2}
\left[\frac{2Et}{\sinh(2Et)}\right]^{\frac{1}{2}}\nonumber\\
&&\times\frac{1}{\beta}\sum_m e^{-Ex^2_m\tanh(Et)}
\ear
where, using the identity (\ref{2c}), the Matsubara sum can be
evaluated with the result
\bear\label{2ah}
\frac{1}{\beta}\sum^\infty_{m=-\infty}e^{-Ex^2_m\tanh Et}
&=&\left[\frac{E}{4\pi \tanh Et}\right]^{\frac{1}{2}}
\sum^\infty_{n=-\infty}e^{-n^2\beta^2E/4\tanh Et}\nonumber\\
&&\times e^{-in(2\pi a+x_1\beta E)}.\ear
Thus we find
\be\label{2ai}
<|\hat\phi(x)|^2>_{sea}=<|\hat\phi(x)|^2>_{T=0}
=\int^\infty_0dte^{-tM^2}(4\pi t)^{-\frac{d-1}{2}}
\frac{E}{4\pi\sinh Et},\ee
\bear\label{2aj}
<|\hat\phi(x)|^2>^{plasma}_{\beta}
&=&\int^\infty_0dte^{-tM^2}(4\pi t)^{-\frac{d-1}{2}}
\frac{E}{4\pi\sinh Et},\\
&&\times \sum^\infty_{n=1}2\cos n(2\pi a+x_1\beta E)
e^{-n^2\beta^2 E/4\tanh Et}\nonumber\ear
One easily verifies that $<|\hat\phi(x)|^2>_{sea}$ coincides
with the corresponding $T=0$ quantity  $<|\hat\phi(x)|^2>_{T=0}$
as it should. This function of course needs UV renormalization.
All dependence on temperature is in $<|\hat\phi(x)|^2>_{\beta}
^{plasma}$. The latter function is finite and it vanishes
exponentially as $T\to 0$. Moreover it is periodic in $x_1$ with
period $\Delta x_1=2\pi/\beta E$, reflecting the equidistant
arrangement of potentials $V_m(x_1)=E^2x^2_m$.

We now proceed to the straightforward calculation of $T_{\mu\nu}^{(\beta)}$.
Using eqs. (\ref{2m})-(\ref{2p}) one easily verifies 
\be\label{2ak}
T_{00}^{(\beta)}(x) = T_{00}^{sea}+T_{00}^{plasma}(x_1)\ee
where 
\be\label{2al}
T_{00}^{sea}=\frac{1}{2}
\int^\infty_0dt e^{-tM^2}(4\pi t)^{-\frac{1}{2}(d-1)}
\frac{E}{4\pi\sinh Et}\left[
\frac{d-1}{2t}+M^2\right],\ee
and
\bear\label{2am}
T_{00}^{plasma}(x_1)&=&
\int^\infty_0dt e^{-tM^2}(4\pi t)^{-\frac{1}{2}(d-1)}
\frac{E}{4\pi\sinh Et}\nonumber\\
&&\times \sum^\infty_{n=1}e^{-n^2\beta^2E/4\tanh Et}\cos n
(2\pi a+x_1\beta E)\nonumber\\
&&\times \left[\frac{d-1}{2t}+M^2+n^2\left(\frac{E\beta}{
2\sinh Et}\right)^2\right].\ear
The energy densities of the virtual sea and thermal plasma
have been obtained by performing the Matsubara sums in eq. (\ref{2ak})
with the help of the identities (\ref{2c}),
(\ref{2t}). For the reader's convenience we give the form in which
the latter identity is used here:
\bear\label{2an}
\frac{E^2}{\beta}\sum^\infty_{m=-\infty}x^2_me^{-x^2_mE\tanh Et}&=&
\left[\frac{E}{4\pi \tanh Et}\right]^{\frac{1}{2}}
\frac{E}{2 \tanh Et}\nonumber\\
&\times&\!\!\!\Big\{\!\!\sum^\infty_{n=-\infty}e^{-n^2\beta^2E/4\tanh Et}
e^{-in(2\pi a+x_1\beta E)}\\
&-&\!\!\!\!\frac{\beta^2E}{\tanh Et}\sum^\infty_{n=1}
n^2 e^{-n^2\beta^2 E/4\tanh Et}\cos n(\beta E x_1 \!\!+\!\! 2\pi a)\Big\}
\nonumber
\ear
Of course, $T^{sea}_{00}$ needs UV renormalization.
In the limit $E\to 0$, $T_{00}^{sea}$ and $T_{00}^{plasma}$ 
above smoothly become the
$E=0$ functions (\ref{2w}), (\ref{2x}).

The other diagonal components of $T_{\mu\nu}^{\ \ (\beta)}$
can be similarly obtained. For brevity we do not write them
down.
Again $T_{11}^{\ \ (\beta)}$ is given by eqs. (\ref{2ak})
-(\ref{2am}) with the signs of the $(d-1)/2t$ and $M^2$
terms reversed. Thus $T^{sea}_{00}=-T^{sea}_{11}$ for any spatial
dimension $d\geq 1$. All nondiagonal components of $T_{\mu\nu}
^{\ \ (\beta)}$ vanish. Interestingly
$T^{sea}_{\mu\nu}$ still vanishes for a massless scalar field in
$d=1$ spatial dimensions. Indeed the tracelessness of $T_{\mu\nu}$
implies $T^{sea}_{00}=T^{sea}_{11}$ and this, combined with
$T^{sea}_{00}=-T^{sea}_{11}$, leads to $T^{sea}_{\mu\nu}=0$.
However, for $d>1$ this tensor does not vanish.

As expected $T^{sea}_{00}$ is independent of position: the uniform
electric field leaves the virtual sea spacially uniform. Physically this seems
reasonable. Virtual particles do not have the prolonged
existence needed to participate in e.g. thermal equilibrium: the sea
remains temperature-independent.

Things are quite different for the thermal plasma. The
particles of the thermal plasma do have prolonged existence,
and they do participate in thermal equilibrium. Also they are
nonuniformly redistributed by the background electric field
$E$ as eq. (\ref{2aj}) shows. The plasma becomes spatially
nonuniform -- in fact periodic as already described with
period $\Delta x_1=2\pi T/E$. As temperature $T\to 0$ with
$E$ fixed, the plasma ceases to exist and  spatial uniformity
is restored. Alternatively, 
holding $T$ fixed and letting
$E\to 0$, the period $\Delta x_1=2\pi/\beta E$ becomes
infinite, thereby restoring spatial uniformity to the plasma. In fact, the total energy contained in the plasma changes discontinuously as we let $E$ tend to zero. Indeed, if we integrate
$T_{00}^{plasma}(x)$ over an integral number of periods
the result is zero. Hence, 
the total energy of the thermal plasma vanishes in the presence
of the uniform electric field. However, in the limit $E\to 0$ 
the nonzero energy density
(\ref{2x}) characterizes the uniform plasma throughout space.

\subsubsection{Constant magnetic field}

To investigate the effect of a uniform background
magnetic field on the virtual sea and
thermal gas of a scalar field it is of
particular interest to consider the case
$d=3$ spatial dimensions. We choose for the
static background potential $A_\mu = (0,0,Bx_1,0)$ leading to the 
magnetic  
field $\vec B = (0,0,B)$. Following our
general notation we then have for the spatial
modes (now $n\to (n, k_2, k_3)$)
\[
\phi_{mnk_2k_3}(\vec x) = \frac{1}{2\pi} e^{i(k_2x_2 + k_3x_3)}  
\varphi_n(x_{k_2})\;,
\]
where $k_2,k_3$ are continuous momentum labels in finite
space and
\[
x_{k_2} = x_1 + \frac{k_2}{B}\;.
\]
 The modes $\varphi_n(x_{k_2})$ are the harmonic oscillator wavefunctions  
(\ref{2ac}) satisfying
\[
\left[ -\partial_1^2 + B^2 x^2_{k_2} \right]\phi_n(x_{k_2}) =  
2B(n+\frac{1}{2}) \phi_n(x_{k_2})\;.
\]
The eigenvalues of $-D^2$ are now
\[
\lambda^2_{mnk_2k_3} = (m\frac{2\pi}{\beta})^2 + 2B(n+\frac{1}{2}) + 
k^2_3\;,
\]
and are independent of $k_2$.

It is already apparent that the uniform magnetic field does not introduce  
spatial non-uniformity into either the virtual sea or the thermal gas. 
Indeed,  
all mode sums involve an integration in  $k_2$ over the infinite interval  
$[-\infty, \infty]$. Since $k_2$ is a continuous variable we can perform 
the  
shift $k_2 \to k_2 - Bx_1$ in the integration variable, thereby absorbing 
the  
$x_1$-dependence into the integration. Thus $T_{\mu\nu}$ and other local  
quantum functions will not depend on $x_1$.

If, however, we compactify the $x_2$ direction perpendicular to the 
magnetic  
field to a circle of perimeter $L$, we are led to a problem
paralleling the one with constant electric field discussed in the 
previous  
subsection. Compact $x_2$ corresponds
to discrete momenta $k_2 = p(\frac{2\pi}{L})$, with $p$ running over all
integers (as in the case of the Matsubara index). The
harmonic oscillator mode equation above  becomes
\[
\left[ -\partial_1^2 + B^2 x^2_p\right]\varphi_n(x_p) =
2B(n+\frac{1}{2})\varphi_n(x_p)\;,
\]
where $x_p = x_1 + p\frac{2\pi}{BL}$.
Again we have an infinite set of harmonic-oscillator
potentials equally spaced at intervals $\Delta x_1 =
\frac{2\pi}{BL}$ along the $x_1$-axis.
Much as in section 2.2.2, local quantities such as
$T_{\mu\nu}$ are periodic in $x_1$ with period
$\Delta x_1$. Clearly one is observing here
something akin to the Quantum Hall Effect.

One could of course extend this discussion to include backgrounds
with both constant electric and magnetic fields.
We shall not persue these interesting matters here,
but rather turn to the case of massless fermions
in 1+1 dimensional space-time.

\section{The finite temperature Schwinger model }
\setcounter{equation}{0}

Electrodynamics of massless fermions in 1+1 dimensions (the
Schwinger Model (SM)) is described by the Lagrangian density
\be\label{Lagrangian}
{\cal L}(x) = -\frac{1}{4}F_{\mu\nu}(x)F^{\mu\nu}(x) +  
{\bar\psi}(x)\left(i\partialslash + e\Aslash\right)\psi(x).
\ee
Our discussion will follow the general lines of the previous section.
Aside from the usual complications of spin,
we shall have to face in the infinite volume limit the existence of an  
infinite number of normalizable zero modes.

\subsection{The case of no electric field}

We choose again (\ref{2q}) for our gauge field configuration.
We emphasize once more that at finite temperature the $A_0$
field cannot be gauged to zero. Hence $A_0 = const$
implies observable effects
\footnote{The same applies to $A_1$ if we were to
compactify space to a circle $S^1$.}. The corresponding
eigenvalue equation for the Dirac field
may be written in the form \footnote{ Our euclidean conventions are 
$\gamma_0 = -\sigma_1, \; , \gamma_1 = \sigma_2 \;, \gamma^5 = \sigma_3 \;$.}
\be\label{eigenvalueequation1}
\left[(i\partial_0 + \frac{2\pi}{\beta}a)
+\gamma^5\partial_1 \right]\psi(x) = \lambda\gamma_0\psi(x)
\ee
with the solution
\be\label{wavefunction1}
\psi_{m,k}(x) = \frac{1}{\sqrt{\beta}}e^{i(2m+1)\frac{\pi}{\beta}x^0}
\left(\begin{array}{cc}
\varphi_k(x_1) \\ \bar \varphi_k(x_1)
\end{array}\right)
\ee
where
\[
\left(\begin{array}{cc}
\varphi_k(x_1) \\ \bar \varphi_k(x_1)
\end{array}\right)
=  \frac{1}{\sqrt{2\pi}}\left(\begin{array}{cc}
A \\ \bar A
\end{array}\right)
e^{ikx_1}
\]
and the eigenvalues
\be\label{Diraceigenvalues}
\lambda_{m}(k) = \pm\sqrt{a_m^2 + k^2}
\ee
where
\[
a_m = -(2m+1)\frac{\pi}{\beta} + \frac{2\pi}{\beta}a
\]
and
\[
\bar A = \mp\sqrt{\frac{a_m-ik}{a_m+ik}}A.
\]
respectively. We normalize the eigenfunctions by choosing
$|A| = \frac{1}{\sqrt{2}}$.
Note that zero modes are absent, unless
$a = (M+\frac{1}{2})$.

Three local quantities are of primary interest to us: The heat kernel 
and corresponding zeta-function densities, as well as the external field
Green function for the construction of the energy-momentum tensor.

\bigskip\noindent
{\it Local heat kernel and zeta-function}

\bigskip
From its definition we have for the matrix valued heat kernel
\bear
h^{(\beta)}(t;x,y)_{\alpha,\beta} &=&
\langle x|{\left[e^{-t(i\Dslash)^2}\right]}
_{\alpha,\beta}|y\rangle\nonumber\\
&=&\sum_m \int \frac{dk}{2\pi}e^{-t{\lambda^2_m(k)}}
{\psi_{m,k} (x)_\alpha \psi_{m,k}^* (y)_\beta}.
\ear
with the corresponding expression for the local zeta function, defined by
\be\label{localzetafunction}
\zeta(s;x,y) = \frac{1}{\Gamma(s)}\int dt\; t^{s-1} h^{(\beta)}(t;x,y)\;.
\ee
Since the eigenvalues come in pairs of positive and negative numbers,
the "chiral" parts of the $2\times 2$ matrix
cancel, leaving us with a diagonal spinor
structure:
\be\label{offheatkernel2}
h^{(\beta)}(t;x,y) = \frac{1}{2\beta}\sum_{m}\int  
\frac{dk}{2\pi}e^{-t\lambda^2_{m}(k)}
e^{i(2m+1)\frac{\pi}{\beta}(x^0-y^0)}
e^{ik(x_1-y_1)}
\left(\begin{array}{cc}1&0\\0&1\end{array}\right).
\ee
We shall be interested only in the diagonal part, $x=y$ of the heat kernel.
Taking the trace we have
\be
\tr h^{(\beta)}(t;x,y) = \frac{1}{\beta}\sum_{m}\int  
\frac{dk}{2\pi}e^{-t\lambda^2_{m}(k)}\; .
\ee
Making use of
the Jacobi identity (\ref{2c}) we may cast $\tr h^{(\beta)}(t;x,x)$
into the form
\be\label{heatkernel2}
\tr h^{(\beta)}(t;x,x) = \tr h(t;x,x)
\left[1 + 2\sum_{m=1}^{\infty} (-1)^m \cos(2\pi m  
a)e^{-m^2\frac{\beta^2}{4t}}\right].
\ee
where $h(t;x,x)$ is the zero-temperature diagonal heat kernel
\be\label{zerotempheat2}
\tr h(t;x,x) = \frac{1}{\sqrt{4\pi t}}\int
\frac{dk}{2\pi} e^{-t k^2} \;.
\ee
Indeed, for $\beta\rightarrow\infty$,
\[
\sum_m \frac{1}{\beta}e^{-t a_m^2} \rightarrow \int \frac{dk^0}{2\pi}
e^{-tk_0^2} = \frac{1}{\sqrt{4\pi t}}\;,
\]
so that we are lead smoothly
from (\ref{heatkernel2}) to (\ref{zerotempheat2})
in the limit $\beta\rightarrow\infty$.

The heat kernel (\ref{heatkernel2}) is interesting in two respects:
i) It exhibits the remarkable periodicity
property in $A_0$ already
encountered in section 2. As commented in that section, this periodicity 
property has its origin in the fact that $A_0(\vec x)$ is 
always gauge equivalent to a configuration taking values in $[0,\frac{2\pi}{\beta}]$. 
ii) 
This exact result provides a counterexample to a claim
\cite{Natividade} made in the literature concerning
the factorization of the temperature dependence
in the thermal heat kernel of a gauged spinor field.

\bigskip
\noindent{\it Energy density}

\bigskip
The energy density is given by
\be\label{DiracEMtensor}
T_{00}(x) = \tr\left[i\gamma_0 D_0(x) G(x,y;A)\right]_{x=y}
\ee
where $G(x,y;A)_{\alpha\beta}$ is the external field Green function
having  spectral representation
\be\label{DiracGreenfunction}
G(x,y;A)_{\alpha\beta} = {\sum_m }\int \frac{dk}{2\pi} 
\;\frac{\psi_{m,k}(x)_\alpha  
\psi_{m,k}^* (y)_\beta}{\lambda_m(k)}
\ee
with the sum extending over the positive and negative eigenvalues  
(\ref{Diraceigenvalues}).
From (\ref{DiracGreenfunction}) and (\ref{wavefunction1}) we obtain for 
the  
Green function (recall that the eigenvalues occur in pairs  
$(\lambda,-\lambda)$, which results in the
off-diagonal structure)
\be\label{Greenfunction1}
G(x,y;A) = \frac{1}{\beta }\sum_{m}\int  
\frac{dk}{2\pi}e^{i(2m+1)\frac{\pi}{\beta}(x_0-y_0)}
e^{ik(x_1-y_1)}
\frac{-1}{|\lambda_{m}(k)|}\left(\begin{array}{cc}
0&\sqrt{\frac{a_m + ik}{a_m - ik}}\\  
\sqrt{\frac{a_m-ik}{a_m+ik}}&0\end{array}\right).
\ee
Performing the differentiation in (\ref{DiracEMtensor}) and combining 
terms,
we obtain
\be\label{Diracenergydensity}
T_{00}(x) = \frac{2}{\beta}\sum_{m}\int \frac{dk}
{2\pi}\frac{a_m^2}{(a_m^2 + k^2)}
\; .
\ee
Notice that this sum diverges and hence requires regularization.
For the special case of a vanishing gauge-potential ($a = 0$) the
result (\ref{Diracenergydensity}) has a familiar interpretation.
Making use of the identity
\[
\sum_{m=-\infty}^{\infty}\frac{1}{(2m+1)^2 + x^2} =
\frac{\pi}{2x}\tanh \frac{\pi x}{2}
\]
we may write (\ref{Diracenergydensity}) for $A_\mu=0$
in the form
\be\label{Diracdistribution}
T_{00}(x) = \frac{2}{\beta}\sum_{m}\int \frac{dk}{2\pi} - \int  
\frac{dk}{2\pi}\omega(k)
+ 2 \int \frac{dk}{2\pi}\omega(k) \frac{1}{e^{\beta\omega(k)} + 1}\;.
\ee
where we have set $|k|=\omega(k)$.
The last term evidently corresponds to the
usal Fermi-distribution for a  gas of massless
fermions, while the second term can be
interpreted as the negative energy contribution
of the Dirac sea. As for the first term, one can identify it with
$\delta^2(0)$, which is independent of $\beta$; hence this term is 
eliminated by temperature-independent renormalization, as should be the case.  

\bigskip
\noindent{\it Effective action and thermal energy}

\bigskip
The effective (euclidean) action is given by
\be\label{effectiveaction}
S_{eff} =  -\frac{1}{2}\ln\;\det\left(i\partialslash + e\Aslash\right)^2 =
 \frac{1}{2}\zeta'(0)
\ee
where $\zeta(s)$ is the zeta-function defined for the spectrum
(\ref{Diraceigenvalues}),
\be\label{zeta}
\zeta(s) = \sum_{-\infty}^{\infty}\int \frac{dk}{2\pi}  
\frac{1}{(\lambda^{2}_{m}(k))^s}
\ee
and where the "prime" on $\zeta(s)$ indicates differentiation
with respect to the argument. With the identification 
$S_{eff} = -ln Z$, with $Z$ the external field partition function,
the average thermodynamic energy is thus given by
\be\label{thermalenergy}
U = \frac{1}{2}\left( \frac{\partial}
{\partial\beta}\zeta'(s)\right)_{s=0}\;.
\ee
In order to allow for a comparison with previous results,
 we first perform the differentiations and then take the limit $s\to 0$ in the 
resulting expression, which then takes the form
\[
\left( \frac{\partial}{\partial\beta}\zeta'(s)\right)_{s=0} =
-2\sum_m\int \frac{dk}{2\pi}\frac{1}{\lambda^2_m(k)}
\frac{\partial}{\partial\beta}\lambda^2_m(k)\;.
\]
It is  now important to observe that our
parametrization of the temporal part of
the gauge field in terms of $A_0 = \frac{2\pi}{\beta}a$
 correctly exhibits the $\beta$-dependence of the  
eigenvalue spectrum. Indeed, as we already remarked above,  $A_0$ may always
be mapped into the interval $[0,\frac{2\pi}{\beta}]$ by an allowed gauge transformation, and correspondingly
the parameter $a$ is taken as lying in the interval
[0,1]. With this
observation we have
\[
\frac{\partial}{\partial\beta}\lambda^2_m(k) =
-\frac{2}{\beta}a_m^2\;,
\]
 so that finally 
\[
U = \frac{1}{2} \frac{\partial}{\partial\beta}\zeta'(0)  
=\frac{2}{\beta}\sum_{m}\int \frac{dk}{2\pi}\frac{a_m^2}{(a_m^2 + k^2)}
\; ,
\]
in agreement with (\ref{Diracenergydensity}). These considerations show
that it is appropriate to define the divergent sum
(\ref{Diracenergydensity}) in terms of the zeta-function via 
(\ref{zeta}).

\subsection{A constant Electric Field}

We now turn to the case of a constant electric field ${\cal E} = -E/e$, 
with the choice
(\ref{2y}) for the gauge potential.
The eigenvalue equation (\ref{eigenvalueequation1}) is now
replaced by
\be\label{eigenvalueequation2}
\left[(i\partial_0 + E x_1 +\frac{2\pi}{\beta}a)
+\gamma^5\partial_1 \right]\psi(x) = \lambda\gamma_0\psi(x)
\ee
Making again the ansatz
\be\label{eigenfunction1}
\psi(x) = \frac{1}{\sqrt{\beta}}e^{i(2m+1)\frac{\pi}{\beta}x_0}
\left(\begin{array}{cc}
\varphi(x_1) \\ \bar \varphi(x_1)
\end{array}\right)
\ee
and defining
\be\label{xm}
x_m = x_1 -(2m+1)\frac{\pi}{E\beta} + \frac{2\pi a}{E\beta}
\ee
we arrive at the coupled set of equations
\bear\label{coupledequations}
\left(E y + \frac{d}{dy}\right)\varphi &=&
-\lambda\bar\varphi\nonumber\\
\left(E y - \frac{d}{dy}\right)\bar\varphi &=& -\lambda\varphi
\ear
where we have set $y = x_m$.
Define the operators
\bear\label{destruction}
a = \frac{1}{\sqrt{2}}\left (\sqrt{| E |} y +  
\frac{1}{\sqrt{|E|}}\frac{d}{dy}\right )\\
a^{\dagger} = \frac{1}{\sqrt{2}}\left ( \sqrt{|E |} y -  
\frac{1}{\sqrt{|E|}}\frac{d}{dy}\right ).
\ear
These operators evidently satisfy the commutation relations of destruction 
and
creation operators, respectively:
\[
[a,a^{\dagger}] = 1
\]
Substituting one equation into the other in (\ref{coupledequations}) we
have, depending on the sign of $E$,
\be\label{alphapositive}
E \; positive:\; \left\{
\begin{array}{cc}
2| E |a a^{\dagger} \bar\varphi = \lambda^2 \bar\varphi\\
2| E |a^{\dagger} a\varphi = \lambda^2 \varphi
\end{array}\right\}
\ee
and
\be\label{alphanegative}
E \; negative:\; \left\{
\begin{array}{cc}
2| E | a^{\dagger}a \bar\varphi = \lambda^2 \bar\varphi\\
2| E | aa^{\dagger} \varphi = \lambda^2 \varphi
\end{array}\right\}
\ee

Now, $2|E|a^\dagger a$ is just the  
Hamiltonian of the harmonic oscillator with the zero-point energy 
ommitted.  
Correspondingly $\varphi$ and $\bar\varphi$ are given by the harmonic 
oscillator  
eigenfunctions.
Defining the ground state $|0>$ by
$a|0> = 0$
we conclude that the eigenstates and corresponding eigenvalues are given by
\[
E \; positive:\; |\Psi^{(\pm)}> =
\left(\begin{array}{cc}
|n>\\ \mp|n-1>
\end{array}\right)\; , \quad \lambda_n = \pm\sqrt{2n|E|}
\]
and
\[
E \; negative :\; |\Psi^{(\pm)}> =
\left(\begin{array}{cc}
|n-1>\\ \mp |n>
\end{array}\right)\; , \quad \lambda_n = \pm\sqrt{2n|E|}
\]
where $|n>$ are the eigenstates of the harmonic oscillator and 
$\lambda^2_n$
are the corresponding energy-eigenvalues without the ``zero-point
energy''.
Denoting by $\varphi_n(x_1)$ the eigenfunctions
of the harmonic oscillator, normalized with respect to the interval
$[-\infty,\infty]$ and setting $y = x_m$, we have for $n\geq 1$ the orthonormalized 
eigenfunctions of the
Dirac operator (\ref{eigenvalueequation2}), 
\be\label{eigenfunctions2a}
\psi_{m,n}^{(\pm)}(x) = 
\frac{1}{\sqrt{2\beta}}e^{i(2m+1)\frac{\pi}{\beta}x_0 }
\left(\begin{array}{cc}
\varphi_n(x_m)\\ \mp \varphi_{n-1}(x_m)
\end{array}\right)\;, n\geq 1\;,
\ee
for positive $E$, and
\be\label{eigenfunctions2}
\psi_{m,n}^{(\pm)}(x) = 
\frac{1}{\sqrt{2\beta}}e^{i(2m+1)\frac{\pi}{\beta}x^0 }
\left(\begin{array}{cc}
\varphi_{n-1}(x_m)\\ \mp \varphi_{n}(x_m)
\end{array}\right)\;, n\geq 1\;,
\ee
for negative $E$, each corresponding to the eigenvalues
\be\label{eigenval2}
\lambda_n = \pm \sqrt{2n|E|}\,,
\ee
respectively.
Since the spectrum corresponds to the absence of the "zero-point energy" 
of  
the harmonic oscillator, we have an infinite set of orthonormalized zero 
modes  
labelled by $m$ and  chirality, of the form
\be\label{zeromodes1}
\phi_{m}^{(+)}(x) = \frac{1}{\sqrt\beta}e^{i(2m+1)\frac{\pi}{\beta}x^0 }
\left(\begin{array}{cc}
\varphi_0(x_m)\\ 0
\end{array}\right)
\ee
for positive $E$, and
\be\label{zeromodes2}
\phi_{m}^{(-)}(x) = \frac{1}{\sqrt\beta}e^{i(2m+1)\frac{\pi}{\beta}x^0 }
\left(\begin{array}{cc}
0 \\ \varphi_{0}(x_m)
\end{array}\right)
\ee
for negative $E$, each corresponding to the eigenvalue $\lambda_0 = 0$.
This is in line with the Atiyah-Singer Index theorem in the infinite
volume limit (see \cite{Sachs}).
Notice that in the case of the zero-modes, the superscript denotes
"chirality".

The wave functions (\ref{zeromodes1}) and (\ref{zeromodes2}) correspond to 
the ground state
wave functions of the harmonic oscillator, localized at the positions
$x_1 = \frac{(2m+1)\pi}{\beta} - \frac{2\pi a}{E\beta}$ with
$m\in Z$. This provides a physical interpretation of the degeneracy of  
the spectrum.
In order to gain a further insight into the problem, we examine next the  
effective Lagrangian giving rise to this degeneracy, as defined in terms 
of  
the "local" $\zeta$-function.

In order to simplify the discussion, we shall restrict ourselves in the  
following to the case where $E$ is positive.

\bigskip
\noindent{\it Effective Lagrangian density}

\bigskip
We begin by considering the local heat kernel.
For the case in question it takes the form
(we now take $E > 0$; we include the zero modes.)
\[
h^{(\beta)}_{\alpha\beta}(t;x,y) = \sum_{m=-\infty}^\infty  
\left(\sum_{n=1}^{\infty}\sum_{\sigma=\pm}
e^{-2nE t}{\psi_{n,m}^{(\sigma)}(x)_{\alpha} {\psi}_{n,m}
^{(\sigma)}(y)_{\beta}^{*}}
+ \phi_m^{(+)}(x)_\alpha \phi_m^{(+)}(y)_\beta^{*}\right)
\]
or explicitly
\bear\label{offdiagonallocalzeta}
h^{(\beta)}_{\alpha\beta}(t;x,y) &=& \sum_{m=-\infty}^{\infty}
\frac{1}{\beta} e^{i(2m+1)\frac{\pi}{\beta}(x_0 - y_0)}\\
&\times &\left(\begin{array}{cc}
\sum_{n=0}^{\infty} e^{-2nE t}{\varphi}_n(x_m ){\varphi}_n^*(y_m )&0\\
0&\sum_{n=1}^{\infty} e^{-2nE t}\varphi_{n-1}(x_m )\varphi_{n-1}^*(y_m )
\end{array}\right)\nonumber
\ear
 The diagonal
matrix structure is again a consequence of the existence of a pair of  
eigenfunctions $\psi^{(\pm)}_n$ corresponding to the eigenvalues  
$\pm\sqrt{2nE}$,
if $n\neq 0$.
We now observe that (note that the sum starts with $n=0$)
\be\label{Defpropagator}
\sum_{n=0}^{\infty} e^{-2nE t}\varphi_n(x_m )\varphi_n^* (y_m )
= e^{E t} <x_m |e^{-tH_{HO}}|y_m >
\ee
where the matrix element on the r.h.s. is the propagation kernel of
the harmonic oscillator known to be given by \footnote{the Hamiltonian in 
our  
case is of the form $H = p^2 + E^2 y^2$, and thus correponds to making 
the  
identifications $m=\frac{1}{2}$, $\omega=2E$ in the conventional
hamiltonian.}
\be\label{propagator}
<x|e^{-tH_{HO} }|y> = \sqrt{\frac{E}{\pi}}\frac{e^{-E t}} {\sqrt{1-e^{-4E 
t}}}
{\rm exp}\left\{ -\frac{E}{2}\frac{(x^2 + y^2)(1 + e^{-4E t})
-4xye^{-2E t}}{1 - e^{-4E t}}\right\}\;.
\ee
Going to the limit of coincident points $x=y$, and taking the trace in
matrix space, we arrive at
\bear\label{ diagonallocalheatkernel }
\tr h^{(\beta)}_{\alpha,\beta}(t;x,x) &=&  \sum_{m=-\infty}^\infty
\frac{1}{\beta}
\left[2\cosh Et <x_m|e^{-tH_{HO} }|y_m>\right]\nonumber\\
&=&\frac{1}{\beta}\sqrt{\frac{E}{\pi}}
\frac{1}{\sqrt{\tanh E t}}
\sum_{m=-\infty}^{\infty}e^{-Ex^2_m \tanh Et}.
\ear
Making use of the identity (\ref{2ah}) we 
may thus write the heat kernel (\ref{ diagonallocalheatkernel }) in the 
form
\[
\tr h^{(\beta)}(t;x,x) = \frac{E}{2\pi}
\left(\frac{1}{\tanh E t}\right)
\left\{1 + 2\sum_{m=1}^{\infty} (-1)^m \cos\left[m(E\beta x_1 + 2\pi  
a)\right]e^{-\frac{m^2\beta^2 E}{4 \tanh E t}}\right\}\;.
\]
In order to compute the effective Lagrangian density we first need to 
subtract the zero-mode contribution:
\[
\tr h'^{(\beta)}(t;x,x) =  \left[\tr h^{(\beta)}(t;x,x) - 
\frac{1}{\beta}\sum_{m=-\infty}^{\infty}  
\varphi_0(x_m)\varphi^*_0(x_m)\right]
\]
where   $\varphi_0(x)$ is the zero-energy harmonic oscillator wave 
function:
\[
\varphi_0(x) = \left(\frac{E}{\pi}\right)^{\frac{1}{4}}
e^{-\frac{E}{2}x^2}\;.
\]
Using (\ref{2c}) we have
\be\label{ heatkder }
\tr h'^{(\beta)}(t;x,x) = \frac{E}{2\pi}\left[f_0(t) + 2\sum_{m=1}^{\infty}  
(-1)^m \cos\left[m(E\beta x_1 + 2\pi a)\right]f_m(t)\right]\;,
\ee
where
\[
f_m(t) = \frac{1}{\tanh E t}e^{-\frac{m^2\beta^2E}{4\tanh E t}}
- e^{-\frac{m^2\beta^2E}{4}}\;.
\]
where the ``prime'' indicates the exclusion of zero-modes. From here we 
obtain  
for the effective Lagrangian \footnote{In the  
$\zeta$-function
regularization the ambiguity in the calculation of
the effective action is well known to be determined
by $\zeta(0)$: $\ln\;\det A = -\zeta'(0) + \zeta(0)\ln\mu^2$, where
$\mu$ is an arbitrary scale parameter.}
\[
{\cal L}_{eff}(x_1) =  
\frac{1}{2}\left[\frac{d}{ds}\zeta^{(\beta)}(s;x,x)\right]_{s=0}
+ \frac{1}{2}\zeta(0;x,x)\ln\mu^2\;,
\]
where
\[
\zeta^{(\beta)}(s;x,x) = \frac{1}{\Gamma(s)}\int_0^{\infty} dt\; t^{s-1}  
h'^{(\beta)}(t;x,x)\;.
\]
and $\mu$ is an arbitrary scale paramenter reflecting the usual ambiguity  
associated with a change in scale of the dimensionful eigenvalues 
$\lambda_n$.  We see that ${\cal L}_{eff}(x_1)$ is again
a periodic function of $x_1$ with period
$\Delta T = \frac{2\pi}{E\beta}$. The degeneracy of the $\lambda_n$ 
spectrum with respect to the Matsubara index labelling the zero modes, 
reflects this fact.
It is clear that correspondingly the energy density will exhibit the same periodicity. Notice that we have again obtained a clean separation of the
$\beta$-independent (sea) and $\beta$-dependent (plasma) contributions.

\bigskip\noindent
{\it Effective action}

\bigskip
Restricting the dimension of our system to
$k$  ``potential wells'' (k zero modes) we have the following relation 
between  
$E\beta$ and the length $L$ of our system:
\be\label{L}
L = k\frac{2\pi}{E\beta}\;.
\ee
Integrating ${\cal L}_{eff}$ over a space-time volume $\beta L$ we thus 
have
for the effective action 
\be
S_{eff} = \frac{1}{2}\left[\zeta'(0) + \zeta(0)\ln\mu^2\right]\;,
\ee
where
\be
\zeta(s) = \int_{0}^{\beta}dx_0\int_{-\frac{L}{2}}^{\frac{L}{2}} dx_1 \zeta^{(\beta)}(s;x,x)\;.
\ee
and the "prime" now means differentiation with respect to $s$. From (\ref{ 
heatkder })  
we have (the cosine term does not contribute to the integral)
\bear
\zeta(s) &=& \frac{k}{\Gamma(s)}\int_0^{\infty} dt\; t^{s-1} 
f_0(t)\nonumber\\
&=&\frac{2k}{\Gamma(s)}\sum_{n=1}^{\infty}\int_0^\infty dt\; t^{s-1} 
e^{-2nE  
t}\nonumber\\
&=&2k\sum_{n=1}^{\infty}\frac{1}{(2nE)^s} =
\frac{2k}{(2E)^s}\zeta_R(s)
\ear
where $\zeta_R(s)$ is the Riemann $\zeta$-function
\[
\zeta_R(s) = \sum_{n=1}^{\infty} \frac{1}{n^s}\; .
\]
Differentiating with respect to $s$, setting $s=0$ and using
$\zeta_R(0) = -\frac{1}{2}$, $\zeta'(0) = -\frac{1}{2}\ln 2\pi$,
as well as (\ref{L}), we obtain for the effective action 
\be\label{Seff}
S_{eff} = -\ln Z = \frac{E\beta L}{4\pi}\ln(\frac{E}{\pi \mu^2})
\ee
in agreement with the result obtained for the corresponding functional determinant on the torus in the presence of a finite number of zero modes
\cite{Sachs}. From (\ref{Seff}) we have for the
average thermal energy
\be\label{energy}
U = -\frac{\partial}{\partial\beta}\ln Z =
\frac{E L}{4\pi}\ln(\frac{E}{\pi \mu^2}).
\ee

Since the energy is temperature independent there is no way of 
normalizing it relative to the $T=0$ case.  
Notice that this temperature  
independence of the total thermal energy is a consequence of the  
spacial periodicity of ${\cal L}_{eff}$, which in turn is the  
result of gauge invariance and periodicity in the time direction.

\section{Conclusion}

We have investigated the effect that uniform background electric and
(in less detail) magnetic fields have on the distribution of
thermal matter fields in unbounded space. By calculating
 the thermal stress tensor $T^{(\beta)}_{\mu\nu}(x)$ 
and effective Lagrangian density  we found that in the presence
of a constant electric field the thermal
plasma distribution of both scalar and fermion fields becomes periodic
along the direction of $\vec E$ with period
$\Delta x = \frac{2\pi T}{E}$, while the virtual sea remains uniform.
On the other hand, a background uniform magnetic field does not lead to a  
spatial non-uniformity in either plasma or sea unless we
compactify one of the spacial directions perpendicular
to the magnetic field (say $x_2$) to a circle. In this
case periodicity of the energy density distribution is
again obtained along a direction $x_1$ perpendicular
to both $x_2$ and to the magnetic field. In both cases
(electric and magnetic field) this periodicity can be traced to the gauge  
invariance of the theory and the possibility of
mapping $A_0(\vec x)$ and $A_2(\vec x)$ into the intervals $[0,\frac{2\pi}{\beta}]$ and  
$[0,\frac{2\pi}{L}]$, respectively,
by a bonafide gauge transformation. 
This periodic structure was shown to reflect an infinite 
degeneracy of the eigenvalue spectrum of the spectral operator.

It is interesting to note that the case of constant magnetic field in 3 
spatial dimensions with one of the spacial dimensions 
in the plane orthogonal to the magnetic 
field compactified, i.e., the quantum Hall problem on a cylinder, is 
formally equivalent to the finite temperature case with constant electric 
field in 1+1 dimensions.  This stems from the fact that at finite 
temperature we are working in 2 dimensional Euclidean space with temporal 
direction compactified.  In the latter case the degeneracy of
the Landau levels in the quantum Hall Effect
corresponds to the number of zero modes of the 
Dirac operator as implied by the Atiyah-Singer Index theorem.

Our exact results for a constant background electromagnetic field have a  
natural extension to arbitrary static background fields
$\vec E(\vec x)$ and $\vec B(\vec x)$. One relevant equation
 to consult is eq. (\ref{2h}). There we see that the space
components of the vector potential are decoupled from the
Matsubara index $m$. For an arbitrary static magnetic field
$\vec B = \nabla \times \vec A$ with $A_0=0$
(recall that at finite temperature $A_0=0$
 cannot be generally achieved by a gauge transformation) one simply makes 
the  replacements $\lambda^2_{mn}
\to (\frac{2\pi m}{\beta})^2 + \omega_n^2$
and $\varphi_{mn} \to \varphi_n$ in eq. (\ref{2y}). The spatial mode equation 
then becomes
\[
\left[ -(\nabla -i\vec A)^2\right]\varphi_n(\vec x) = \omega_n^2 \varphi_n(\vec 
x)\;.
\]
For arbitrary $\vec B(\vec x)$ one thus has a situation much like  the  
non-gauge theory of section 2.1, leading to some non-uniform distribution
of both the sea and plasma components comparable
to what physical boundaries would cause.

The situation is quite different when $A_0$ is non-zero.
Then the static background field affects the sea
and plasma quite differently. While
plasma periodicity along the direction of
$\vec E$ is strictly true only for constant
$\vec E$, one would expect for any electric
field  which is only weakly dependent
on $\vec x$ a roughly periodic response from
the plasma, and a nearly uniform distribution of the sea-component.
Mathematically
we have an infinite set of eigenvalue
equations with a \underbar{different} Schr\"odinger
like background potential
$V_m(\vec x) = [A_0(\vec x) - 2\pi m/\beta]^2$
 for each Matsubara frequency. Note that
 $m\to m+N$ corresponds to performing an allowed
gauge transformation with gauge function $\lambda = x_0(2\pi N/\beta)$,
and that $V_{m+N}(\vec x)$ and $V_{m}(\vec x)$
are connected by this gauge transformation.
This situation differs fundamentally from the non-gauge case,
where the potential $V(\vec x)$ does not bear the label $m$. Since Green  
functions, the energy momentum tensor, etc., are given in terms of equally  
weighted sums over all the individual
problems labelled by $m$, they are explicitely gauge invariant.

In general the diagonal heat kernel of a scalar or
fermion quantum field at finite temperature $T>0$
in a {\it static} background is expected to factorize
in the following way:
\[
h^{(\beta)}(t;x,x) = h(t;x,x)_{T=0} \left[1 + f(t;x;T)\right]
\]
where $h(t;x,x)_{T=0}$ is the temperature-zero heat kernel for the same  
background, and $f(t,x,T)$ is some function of the temperature $T$, the  
diffusion or ``proper'' time $t$ and the spatial position
$\vec x$. This function $f$ vanishes exponentially as
either $T\to 0$ or $t \to 0$. The factorization above
is motivated by the expectation that $h^{(\beta)}(t;x,x)$ separates quite  
generally for a static background into an UV divergent sea part, and an 
UV  
finite gas part
\[
h^{(\beta)}(t;x,x) = h(t;x,x)_{sea} + h(t;\vec x)_{gas}
\]
where
\[
h(t;x,x)_{sea} = h(t;x,x)_{T=0}\;.
\]
Defining $f(t;x;T)$ by
\[
h(t;\vec x)_{gas} = f(t;x;T)h(t;x,x)_{T=0}
\]
we arrive at the factorization above. Known properties of $h(t;\vec 
x)_{gas}$
lead to the stated properties of $f$.

It is a matter of some interest to study the function $f(t,\vec x;T)$.
Let us list the explicit examples computed in this paper.

Nongauge scalar theory:
\[
1+f=\sum^\infty_{n=-\infty}e^{-n^2\beta^2/4t}.
\]
Gauged scalar theory with $A_0=Ex_1+2\pi a/\beta$:
\[
1+f=\sum^\infty_{n=-\infty}e^{-n^2\beta^2E/4\tanh Et}
e^{-in\beta A_0}.
\]
Schwinger model with $A_0=Ex_1+2\pi a/\beta$:
\[
1+f=\sum^\infty_{n=-\infty}(-)^ne^{-n\beta^2E/4\tanh Et}
e^{-in\beta A_0}.\]
From the gauge theory examples we see that $f(t,\vec x;T)$
in general depends on $A_0$ when a gauge potential background
is involved.

In ref. \cite{Natividade} it is argued that $f$ has the simple form
\[
1+f=\sum^\infty_{n=-\infty}(-)^ne^{-n^2\beta^2/4t}
\]
for a gauged spinor theory in a general (even time-dependent)
background gauge potential $A_\mu$. This claim is incorrect: the result 
for the Schwinger model provides a simple, explicit counterexample.

\section*{Acknowledgements}
One of the authors (AAA) thanks R. Viollier and 
the theory group of the  Department of Physics, University of Capetown,
for their kind hospitality during part of his 1997/98 sabbatical.
Support from Penn state (RDG and Lehigh Valley) is gratefully
acknowledged. Another  of the authors (KDR) thanks the Physics Department,
University of Stellenbosch, for their kind hospitality.
This work was supported by a grant from the Foundation
of Research Development of South Africa.

\newpage


\end{document}